\documentclass[aps,preprint,groupedaddress,showpacs,longbibliography,nofootinbib]{revtex4-2}
\usepackage{amssymb}
\usepackage{amsmath}
\usepackage[breaklinks,colorlinks,citecolor=green,urlcolor=blue,linkcolor=red]{hyperref}
\usepackage{graphicx}
\usepackage{epstopdf}
\usepackage{diagbox}
\usepackage{float}
\usepackage{subfigure}
\usepackage{multirow}
\usepackage{CJK}
\usepackage{tikz}
\usepackage{enumerate}

\begin{document}



\title{\boldmath $C\!P$ asymmetries corresponding to the imaginary parts of the interference terms in cascade decays of heavy hadrons}


\author{Jing-Juan Qi}
\email{qijj@mail.bnu.edu.cn}
\affiliation{College of Information and Intelligence Engineering, Zhejiang Wanli University, Ningbo, Zhejiang 315101, China}

\author{Jian-Yu Yang}
\email{yangjy@stu.usc.edu.cn}
\affiliation{School of Nuclear Science and Technology, University of South China, Hengyang, Hunan 421001, China.}

\author{Zhen-Hua Zhang}
\email[corresponding author, ]{zhangzh@usc.edu.cn}
\affiliation{School of Nuclear Science and Technology, University of South China, Hengyang, Hunan 421001, China.}
\date{\today}


\date{\today}

\begin{abstract}
A mechanism of generating CP violation through the imaginary part of the interference of two amplitudes is proposed.
This mechanism has shown clear evidence in decays such as $B^\pm\to \pi^\pm\pi^+\pi^-$.
The proposed mechanism is helpful in searching for CP violation in bottom and charmed baryon decay processes.
\end{abstract}
\maketitle

CP violation (CPV) plays an important role in the Standard Model of particle physics and cosmology.
It has been observed in the decays of $K$, $B$, and $D$ mesons \cite{Christenson:1964fg,LHCb:2019hro,BaBar:2001ags,Belle:2001zzw,LHCb:2013syl,Workman:2022ynf}.
On the contrary, CPV has not been observed in any baryon decay process, which deserves further theoretical and experimental study.

CPV induced by time-reversal-odd (T-odd) correlations plays a special role in searching for CPV in heavy hadron decays \cite{Donoghue:1987wu,Valencia:1988it},  
among which those induced by the triple-product asymmetries (TPAs), i.e., the distribution asymmetry corresponding to the signature of the triple-products of the momenta and/or spins of the particle involved, are the most extensively studied ones \cite{Valencia:1988it,Dunietz:1990cj,Golowich:1988ig,Kayser:1989vw,Bensalem:2000hq,Bensalem:2002pz,Bensalem:2002ys,Durieux:2015zwa,Gronau:2015gha,Shi:2019vus,Wang:2022fih}. 
TPA induced CP asymmetries (CPAs) have been investigated experimentally in decay channels such as: $D^0\to K^+ K^- \pi^+\pi^-$, $D^+\to K^+ K^- \pi^+\pi^0$, $D_{(s)}^{+}\to K^{+} K^{-}\pi^{+}\pi^{0}$, $D_{(s)}^{+}\to K^{+} \pi^{-}\pi^{+}\pi^{0}$, $D^0 \to K_S^0 K_S^0 \pi^+ \pi^-$, $\Lambda^{0}_{b}\to pK^{-}\pi^{+}\pi^{-}$, $\Lambda^{0}_{b}\to pK^{-}K^{+}K^{-}$ and $\Xi^{0}_{b}\to pK^{-}K^{-}\pi^{+}$ \cite{Belle:2005qtf,BaBar:2010xrb,LHCb:2014djq,LHCb:2016yco,LHCb:2016qbq,Belle:2017zvp,LHCb:2018fpt,Belle:2018pcz,LHCb:2018mzv,LHCb:2019oke,Belle:2023bzn}.
Although a non-zero TPA was observed in the decay channel $\Lambda^{0}_{b}\to p\pi^{-}\pi^{+}\pi^{-}$ \cite{LHCb:2019oke}: 
$a_{P}^{\hat{T}-\text{odd}}=(-4.0\pm0.7\pm0.2)\%$,
TPA induces CP asymmetry (TP-CPA) has never been observed yet.

It happens that TPAs usually get contributions from the imaginary part of some interfering terms of two amplitudes.
It can be shown that TP-CPA is usually proportional to the cosine of the relative strong phases: 
\begin{equation}
  \Im (a b^\ast)-\Im (\overline{a} \overline{b}^\ast)\sim\sin\phi\cos\delta,
\end{equation}
where $a$ and $b$ are the two amplitudes, $\phi$ is the weak phase between them, while $\delta$ is the strong phase difference,  $\overline{a}$ and $\overline{b}$ are respectively the CP-conjugates of $a$ and $b$.
In this short paper, we propose another mechanism which can generate CPAs that are also proportional to the imaginary part of some interfering terms in multi-body decays of heavy hadrons, just as TP-CPA does.
In this mechanism, the interference occurs between the amplitudes corresponding to two intermediated resonances.
Unlike the situation of TP-CPA which has never been observed experimentally, we will show that the CPV induced by the currently proposed mechanism has in fact already left clear evidence in three-body decays of bottom mesons such as  $B^\pm\to \pi^\pm\pi^+\pi^-$.


Let's take a three-body decay $\mathbb{H}\to  abc$  to illustrate, where $\mathbb{H}$ is the mother hadron, and $a$, $b$, and $c$ are the three hadrons in the final state.
We will consider cascade decays of the form $\mathbb{H}\to r c$, $r\to ab$, where $r$ represents an intermediate resonance. 
The decay amplitudes can be expressed in the form of
\begin{equation}
  \mathcal{M}=\sum_r\frac{\mathcal{A}_r}{s_r},
\end{equation}
where $\mathcal{A}_r$ is the decay amplitude for the cascade process $\mathbb{H}\to r c$, $r\to ab$,  $s_r$ is the reciprocal of the Breit-Wigner factor, which approximately takes the from
$s_r = s-m_r^2+\mathrm{i}m_r\Gamma_r$, with $s$ the invariant mass of the $ab$ system, the summation over $r$ indicates that there can be more than one intermediate resonances.
The decay amplitude squared can then be expressed as
\begin{equation}
  \overline{\left|\mathcal{M}\right|^2}=\sum_{rr'}\frac{\mathcal{A}_r\mathcal{A}_{r'}^\ast}{s_rs_{r'}^\ast},
\end{equation}
where the overline above the decay amplitude squared indicates that there may be spin or helicity summation over the initial and the final states. Note that the possible spin or helicity summations have been omitted on the right hand side of the above equation.
One can see that there can be interferences between different intermediate resonances, provided that the decay amplitude is dominated by at least two intermediate resonances in some regions of the phase space.
This happens when the masses of the intermediate resonances are close to each other, and their decay widths are not small.
We consider the simplest situation when two intermediate resonances dominate in some regions of the phase space. 
In the interfering region of the phase space, the decay amplitude squared can then be expressed as
\begin{eqnarray}\label{eq:Msq}
  \overline{\left|\mathcal{M}\right|^2}\approx\frac{\left|\mathcal{A}_1\right|^2}{\left|s_1\right|^2}+\frac{\left|\mathcal{A}_2\right|^2}{\left|s_2\right|^2} +2\Re\left(\frac{\mathcal{A}_1\mathcal{A}_{2}^\ast}{s_1s_{2}^\ast}\right),
\end{eqnarray}
For the convenience of the subsequent analysis, we can further expressed the interfering term as 
\begin{equation}\label{eq:Interf}
 \Re\left(\frac{\mathcal{A}_1\mathcal{A}_{2}^\ast}{s_1s_{2}^\ast}\right)= \frac{\Re\left({\mathcal{A}_1\mathcal{A}_{2}^\ast}\right)\Re\left({s_1 s_{2}^\ast}\right)+\Im\left({\mathcal{A}_1\mathcal{A}_{2}^\ast}\right)\Im\left({s_1 s_{2}^\ast}\right)}{\left|s_1 s_2\right|^2}.
\end{equation}
Note that  we have devided the two-resonance-interfering term into two different terms, which respectively proportional to $\Re\left({s_1 s_{2}^\ast}\right)=m_1\Gamma_1 m_2\Gamma_2+\left(s-m_1^2\right)\left(s-m_2^2\right)$ and $\Im\left({s_1 s_{2}^\ast}\right)=\left(s-m_2^2\right)m_1\Gamma_1-\left(s-m_1^2\right) m_2\Gamma_2$.
The second term on the right hand side of Eq. (\ref{eq:Interf}) is especially interesting since it is proportional to the imaginary part of the interfering term $\mathcal{A}_1\mathcal{A}_{2}^\ast$.
By generally expressing $\mathcal{A}_1$ and $\mathcal{A}_{2}$ in terms of amplitudes with different weak phases, $\mathcal{A}_1=\sum_k  \mathcal{A}_1^k$ and $\mathcal{A}_{2}=\sum_{k'}  \mathcal{A}_2^{k'}$, it can be seen that  CPA induced by the imaginary part of the interfering term $\mathcal{A}_1\mathcal{A}_{2}^\ast$ will be a collection of terms which are proportional to the cosine of some unitary phases:
\begin{equation}
 A_{\Im \left({\mathcal{A}_1\mathcal{A}_{2}^\ast}\right)}\equiv
 \Im \left({\mathcal{A}_1\mathcal{A}_{2}^\ast}\right)-\Im\left({\overline{\mathcal{A}_1}~\overline{\mathcal{A}_{2}}^\ast}\right)\propto\sum_{kk'} \left|\mathcal{A}_1^{k} \mathcal{A}_2^{k'}\right|\sin\phi_{kk'}\cos\delta_{kk'},
\end{equation}
where $\overline{\mathcal{A}_1}$ and $\overline{\mathcal{A}_{2}}$ are the corresponding amplitudes for the CP-conjugate process, 
 $\phi_{kk'}$ and $\delta_{kk'}$ are respectively the weak and unitary phase differences between $\mathcal{A}_1^{k}$ and $\mathcal{A}_{2}^{k'}$. 
As comparison, CPA induce by the real part of the interfering term $\mathcal{A}_1\mathcal{A}_{2}^\ast$ takes the form 
\begin{equation}
A_{\Re \left({\mathcal{A}_1\mathcal{A}_{2}^\ast}\right)}\equiv
\Re \left({\mathcal{A}_1\mathcal{A}_{2}^\ast}\right)-\Re\left({\overline{\mathcal{A}_1}~\overline{\mathcal{A}_{2}}^\ast}\right)\propto\sum_{kk'} \left|\mathcal{A}_1^{k}\mathcal{A}_2^{k'}\right|\sin\phi_{kk'}\sin\delta_{kk'}.
\end{equation}

It is more common that the decay width of one of the resonances is considerably larger than that of the other one.
For the current situation, suppose that $\Gamma_1 >\Gamma_2$. 
We also expect that the masses of the two resonances are close to each other, so that we have $\Gamma_1>m_1-m_2$.
Besides, since we are focusing on the interfering effect, it will be enough for us to constrain $s$ in the interfering region, so that we have $\Gamma_1>|s-m_1^2|/m_1$ and $\Gamma_1> |s-m_2^2|/m_2$.
With the above consideration, one can see that $\Re\left({s_1 s_{2}^\ast}\right)$ is dominated by the first term, so that one has
$\Re\left({s_1 s_{2}^\ast}\right)\approx m_1\Gamma_1 m_2\Gamma_2$.
Meanwhile, we can re-arrange $\Im\left({s_1 s_{2}^\ast}\right)$ into the form 
\begin{equation}\label{eq:ims1s2}
  \Im\left({s_1 s_{2}^\ast}\right)=m_1\Gamma_1(1 -\frac{m_2\Gamma_2}{m_1\Gamma_1})\left(s-{m_2^\prime}^2\right),
\end{equation}
 where ${m_2^\prime}=m_2\sqrt{(1-\frac{m_1\Gamma_2}{m_2\Gamma_1})/(1 -\frac{m_2\Gamma_2}{m_1\Gamma_1})}$.
Now the difference between the behaviour of $\Re\left({s_1 s_{2}^\ast}\right)$ and $\Im\left({s_1 s_{2}^\ast}\right)$ is obvious: while the latter tends to change sign as $s$ passes through ${m_2^{\prime}}^2$, the former does not.

The aforementioned behaviour difference allows us to isolate the contributions of the term $\Im \left({\mathcal{A}_1\mathcal{A}_{2}^\ast}\right)$ to CPV. 
For a CPV observable which is defined on the phase space region ${m_2^\prime}^2-\Delta_-<s<{m_2^\prime}^2+\Delta_+$, with $\Delta_-$ and $\Delta_+$ about the same order as ${\Gamma_2}^2$ and/or $(m_1-m_2)^2$,
\begin{equation}
  A_{CP}\equiv \frac{\int_{{m_2^\prime}^2-\Delta_-}^{{m_2^\prime}^2+\Delta_+} \left(\overline{\left|\mathcal{M}\right|^2}-\overline{\left|\overline{\mathcal{M}}\right|^2}\right)  d s}{\int_{{m_2^\prime}^2-\Delta_-}^{{m_2^\prime}^2+\Delta_+}  \left(\overline{\left|\mathcal{M}\right|^2}+\overline{\left|\overline{\mathcal{M}}\right|^2}\right)  d s},
\end{equation}
where $\overline{\mathcal{M}}$ is the decay amplitude for the CP-conjugate process,
the contribution of the term  $\Im \left({\mathcal{A}_1\mathcal{A}_{2}^\ast}\right)$ to  $A_{CP}$ is pretty much negligible, because that there is strong cancellation for the term $\Im\left({s_1 s_{2}^\ast}\right)$ when integrating over ${m_2^\prime}^2-\Delta_-<s<{{m_2^\prime}^2}$ and ${{m_2^\prime}^2}<s<{m_2^\prime}^2+\Delta_+$.
To avoid this cancellation,
one can introduce a sign factor in the definition of the above CPV observables:
\begin{equation}\label{eq:ImACP}
  A_{CP}^{\Im}\equiv \frac{\int_{{m_2^\prime}^2-\Delta_-}^{{m_2^\prime}^2+\Delta_+} \left(\overline{\left|\mathcal{M}\right|^2}-\overline{\left|\overline{\mathcal{M}}\right|^2}\right) ~ \text{sgn}\! \left(s-{m_2^\prime}^2\right) d s}{\int_{{m_2^\prime}^2-\Delta_-}^{{m_2^\prime}^2+\Delta_+} \left(\overline{\left|\mathcal{M}\right|^2}+\overline{\left|\overline{\mathcal{M}}\right|^2}\right)  d s},
\end{equation}
where the sign function $~\text{sgn}$ is defined as 
\begin{equation}
  ~\text{sgn} (x)=\left\{
                            \begin{array}{cc}
                              +1, & x>0; \\
                              -1, & x<0. \\
                            \end{array}
  \right.
\end{equation}

It can be shown that the newly introduced CPV observable $A_{CP}^{\Im}$ gets its contribution mainly from the term  $\Im \left({\mathcal{A}_1\mathcal{A}_{2}^\ast}\right)$.
To see this, one need to re-express Eq. (\ref{eq:ImACP}) as
\begin{equation}\label{eq:ImACPexp}
  A_{CP}^{\Im} \propto A_{\Im\left({\mathcal{A}_1\mathcal{A}_{2}^\ast}\right)} I_1 ^{(\Delta_-,\Delta_+)}
  +A_{\Re\left({\mathcal{A}_1\mathcal{A}_{2}^\ast}\right)} I_2 ^{(\Delta_-,\Delta_+)}
  +A_{\left|\mathcal{A}_1\right|^2} I_3^{(\Delta_-,\Delta_+)} +A_{\left|\mathcal{A}_2\right|^2}I_4^{(\Delta_-,\Delta_+)},
\end{equation}
where $A_{\left|\mathcal{A}_1\right|^2}\equiv \left|\mathcal{A}_1\right|^2-\left|\overline{\mathcal{A}_1}\right|^2$ and $A_{\left|\mathcal{A}_2\right|^2}\equiv \left|\mathcal{A}_2\right|^2-\left|\overline{\mathcal{A}_2}\right|^2$ are respectively measures of CPV in the two-body decays $\mathbb{H}\to 1 c$ and $\mathbb{H}\to 2 c$,
and the four integrals are respectively defined as
\begin{align}
  I_1 ^{(\Delta_-,\Delta_+)} &\equiv m_1\Gamma_1(1 -\frac{m_2\Gamma_2}{m_1\Gamma_1})\int_{{m_2^\prime}^2-\Delta_-}^{{m_2^\prime}^2+\Delta_+} \frac{ \left(s-{m_2^\prime}^2\right)\text{sgn}\! \left(s-{m_2^\prime}^2\right) }{\left|s_1s_2\right|^2} d s,\\
  I_2 ^{(\Delta_-,\Delta_+)} &\equiv 
  \int_{{m_2^\prime}^2-\Delta_-}^{{m_2^\prime}^2+\Delta_+} \frac{\left[m_1\Gamma_1 m_2\Gamma_2+\left(s-m_1^2\right)\left(s-m_2^2\right)\right]\text{sgn}\! \left(s-{m_2^\prime}^2\right) }{\left|s_1s_2\right|^2} d s,\\
  I_3 ^{(\Delta_-,\Delta_+)} &\equiv 
  \int_{{m_2^\prime}^2-\Delta_-}^{{m_2^\prime}^2+\Delta_+} \frac{\text{sgn}\! \left(s-{m_2^\prime}^2\right) }{\left|s_1\right|^2} d s,\\
  I_4 ^{(\Delta_-,\Delta_+)} &\equiv 
  \int_{{m_2^\prime}^2-\Delta_-}^{{m_2^\prime}^2+\Delta_+} \frac{\text{sgn}\! \left(s-{m_2^\prime}^2\right) }{\left|s_2\right|^2} d s.
\end{align}
As can be seen from the above equations, the presence of the factor $\text{sgn}\! \left(s-{m_2^\prime}^2\right)$ makes sure that there are cancellations for the integrals between ${m_2^\prime}^2-\Delta_-<s<{m_2^\prime}^2$ and ${m_2^\prime}^2<s<{m_2^\prime}^2+\Delta_+$ parts in $I_2^{(\Delta_-,\Delta_+)}$, $I_3^{(\Delta_-,\Delta_+)}$, and $I_4^{(\Delta_-,\Delta_+)}$. 
In contrast, it is the same factor, together with $\left(s-{m_2^\prime}^2\right)$, guarantees that there are no such kind of cancellation in $I_1^{(\Delta_-,\Delta_+)}$.
Consequently, the contribution of the first term on the right hand side of Eq (\ref{eq:ImACPexp}) containing $\Im \left({\mathcal{A}_1\mathcal{A}_{2}^\ast}\right)$ is amplified, while the last three terms are attenuated. 

Clearly, Eq. (\ref{eq:ImACP}) can not isolate the contribution of the term $\Im \left({\mathcal{A}_1\mathcal{A}_{2}^\ast}\right)$ in an ideal way,
in the sense that the introduction of the sign factor can not entirely eliminate the contributions of other terms to $A_{CP}^{\Im}$. 
Despite that, it does not affect the fact that $A_{CP}^{\Im}$ gets its main contribution from the term  $\Im \left({\mathcal{A}_1\mathcal{A}_{2}^\ast}\right)$.
It is possible to further suppress the contribution of other terms in Eq. (\ref{eq:ImACPexp}).
For example,  one can arrange the values of $\Delta_-$ and $\Delta_+$ to fulfill the condition that $I_2 ^{(\Delta_-,\Delta_+)}=0$, $I_3 ^{(\Delta_-,\Delta_+)}=0$, or $I_4 ^{(\Delta_-,\Delta_+)}=0$.
However, it does not seem necessary to do so, because $A_{CP}^{\Im}$ is worth to be measured especially when one can not find CPV elsewhere, which means that $A_{\Re\left({\mathcal{A}_1\mathcal{A}_{2}^\ast}\right)}$, $A_{\left|\mathcal{A}_1\right|^2}$, and $A_{\left|\mathcal{A}_2\right|^2}$ are likely to be small. 
Consequently, it is unnecessary to arrange the values of $\Delta_-$ and $\Delta_+$. One can even just simply choose $\Delta_-=\Delta_+$. 

This seems to be a good place to stress that although we adopt the Breit-Wigner formula for the propagator factors, the basics idea does not depends on this choice.
Explicitly, the sign changing characteristics in Eq. (\ref{eq:ims1s2})  is independent on the choise of the propagator factor formula. 
However, the value of the zero point ${m_2^\prime}^2$ does depend on the choice, making the value of ${m_2^\prime}^2$ may be different in different choice of propagator factor formulas.
To reduce the uncertainties caused by the values of the zero point ${m_2^\prime}^2$, an alternative choice for Eq. (\ref{eq:ImACP}) is to remove a small section in the integral: 
\begin{equation}\label{eq:ImACPalt}
  A_{CP}^{\Im,\text{alt}}\equiv \frac{\left(\int_{{m_{2-}^\prime}^2-\Delta_-}^{{m_{2-}^\prime}^2} -\int_{{m_{2+}^\prime}^2}^{{m_{2+}^\prime}^2+\Delta_+}\right) \left(\overline{\left|\mathcal{M}\right|^2}-\overline{\left|\overline{\mathcal{M}}\right|^2}\right)  d s}{\left(\int_{{m_{2-}^\prime}^2-\Delta_-}^{{m_{2-}^\prime}^2} +\int_{{m_{2+}^\prime}^2}^{{m_{2+}^\prime}^2+\Delta_+}\right) \left(\overline{\left|\mathcal{M}\right|^2}+\overline{\left|\overline{\mathcal{M}}\right|^2}\right)  d s},
\end{equation}
where $m_{2-}^\prime$ and $m_{2+}^\prime$ are chosen to fulfill the condition  $m_{2-}^\prime< m_{2}^\prime <m_{2+}^\prime$.

It often happens that the quantum number of the two resonances are different, which means that the kinematical behaviour of the interfering terms is different from the non-interfering terms.
For example, the interference of a scalar resonance and a vector resonance leads to forward-backward asymmetry (FBA) in the decay-angular distributions, which can further induce corresponding CPAs.
In the phase space region ${m_2^\prime}^2-\Delta_-<s<{m_2^\prime}^2\textcolor{red}{+}\Delta_+$, the FBA induced CPA \footnote{The FB-CPA defined in Eq. (\ref{eq:FBCPA}) is actually called direct-CPA-subtracted FB-CPA in Ref. \cite{Wei:2022zuf}.}
can be defined as \cite{Zhang:2021zhr,Zhang:2022emj,Wei:2022zuf}
\begin{equation}\label{eq:FBCPA}
  A_{CP}^{FB}\equiv  \frac{\int_{{m_2^\prime}^2-\Delta_-}^{{m_2^\prime}^2+\Delta_+}  \int_{-1}^{+1} \left(\overline{\left|\mathcal{M}\right|^2}-\overline{\left|\overline{\mathcal{M}}\right|^2}\right)  ~{\text{sgn}}\! \left(c_{\theta}\right)  d c_{\theta} d s}{\int_{{m_2^\prime}^2-\Delta_-}^{{m_2^\prime}^2+\Delta_+}  \int_{-1}^{+1} \left(\overline{\left|\mathcal{M}\right|^2}+\overline{\left|\overline{\mathcal{M}}\right|^2}\right)  d c_{\theta} d s},
\end{equation}
where $c_{\theta}\equiv\cos\theta$, with $\theta$ being a helicity angle which is defined as the angle between momenta of $a$ and $c$ in the centre-of-mass frame of the $ab$ system.
It can be easily shown that this $A_{CP}^{FB}$ gets the contribution mainly from $ \Re \left({\mathcal{A}_1\mathcal{A}_{2}^\ast}\right)$.
Just as the above, by introducing the sign factor $\text{sgn}\! \left(s-{m_2^\prime}^2\right)$,
one can obtain a CPV observable which gets its contribution mainly from the term $ \Im \left({\mathcal{A}_1\mathcal{A}_{2}^\ast}\right)$:
\begin{equation}\label{eq:FBCPAIm}
  A_{CP}^{FB,\Im}\equiv  \frac{\int_{{m_2^\prime}^2-\Delta_-}^{{m_2^\prime}^2+\Delta_+}  \int_{-1}^{+1} \left(\overline{\left|\mathcal{M}\right|^2}-\overline{\left|\overline{\mathcal{M}}\right|^2}\right)  ~\text{sign}\! \left(c_{\theta}\right) ~\text{sgn}\! \left(s-{m_2^\prime}^2\right) d c_{\theta} d s}{\int_{{m_2^\prime}^2-\Delta_-}^{{m_2^\prime}^2+\Delta_+}  \int_{-1}^{+1} \left(\overline{\left|\mathcal{M}\right|^2}+\overline{\left|\overline{\mathcal{M}}\right|^2}\right)  d c_{\theta} d s}.
\end{equation}

In general, it is also possible that the interfering terms enter in the angular distributions of the final particles.
For example, for the partial-wave CPA which can be defined as \cite{Zhao:2024ren,Yu:2024cjd}
\begin{equation}
  A_{CP}^{(l)}\equiv  \frac{\int_{{m_2^\prime}^2-\Delta_-}^{{m_2^\prime}^2+\Delta_+}  \int_{-1}^{+1} \left(\overline{\left|\mathcal{M}\right|^2}-\overline{\left|\overline{\mathcal{M}}\right|^2}\right)  ~P_l\! \left(c_{\theta}\right)  d c_{\theta} d s}{\int_{{m_2^\prime}^2-\Delta_-}^{{m_2^\prime}^2+\Delta_+}  \int_{-1}^{+1} \left(\overline{\left|\mathcal{M}\right|^2}+\overline{\left|\overline{\mathcal{M}}\right|^2}\right)  d c_{\theta} d s},
\end{equation}
with $P_l$ the $l$-Legendre polynomial,
it gets the contribution from the real parts of certain interfering terms.
The contribution of the imaginary parts of these interfering terms to CPV are mainly contained in
\begin{equation}
  A_{CP}^{(l),\Im}\equiv  \frac{\int_{{m_2^\prime}^2-\Delta_-}^{{m_2^\prime}^2+\Delta_+}  \int_{-1}^{+1} \left(\overline{\left|\mathcal{M}\right|^2}-\overline{\left|\overline{\mathcal{M}}\right|^2}\right) ~P_l\! \left(c_{\theta}\right) ~\text{sgn}\! \left(s-{m_2^\prime}^2\right) d c_{\theta} d s}{\int_{{m_2^\prime}^2-\Delta_-}^{{m_2^\prime}^2+\Delta_+}  \int_{-1}^{+1} \left(\overline{\left|\mathcal{M}\right|^2}+\overline{\left|\overline{\mathcal{M}}\right|^2}\right)  d c_{\theta} d s}.
\end{equation}

\begin{figure}[H]
    \centering  
        \includegraphics[width=.9\linewidth]{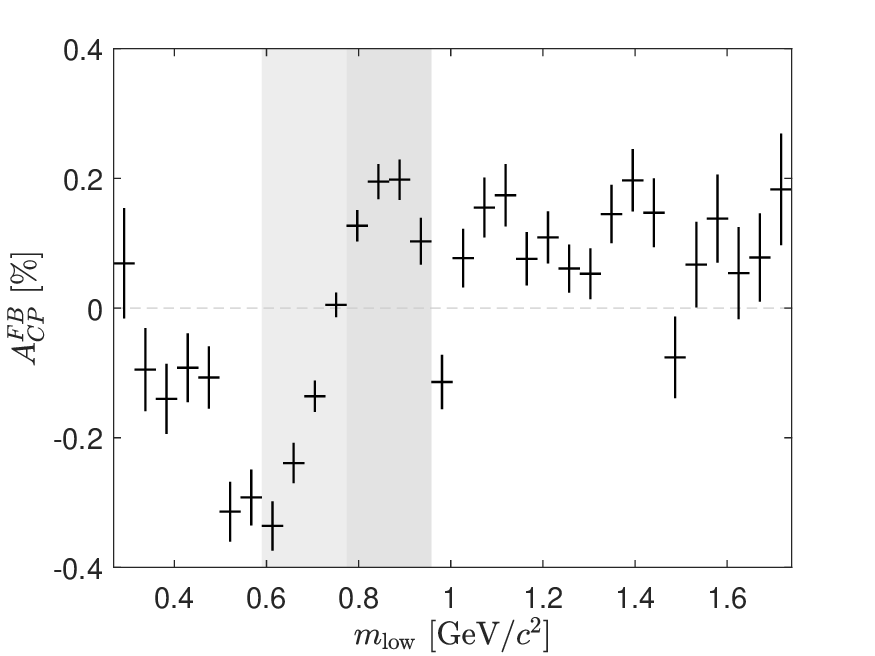}
    \caption{The FB-CPA of each bin distributed in the low $m_{\text{low}}$ region for $B^\pm\to\pi^+\pi^-\pi^\pm$.
    To obtain the FB-CPAs of each bin, both the event yields differences $N_{B^-}-N_{B^+}$ and the total event yields combined $N_{B^-}+N_{B^+}$ are essential. The former can be directly read off from Fig. 12 of Ref. \cite{LHCb:2019sus}, while the later can be obtained according to $N_{B^-}+N_{B^+}=\sigma^2$, where $\sigma$ is the error of $N_{B^-}-N_{B^+}$ and can be read off from the same figure. Note that we have simply assumed that $\sigma$ contains only the statistical error.}
    \label{fig:FBCPA3pi}
\end{figure}

Unlike the TP-CPAs, CPA induced by the imaginary part of the interfering term $\Im \left({\mathcal{A}_1\mathcal{A}_{2}^\ast}\right)$ has in fact been observed experimentally in a sense.
Take the decay $B^\pm\to \pi^\pm\pi^+\pi^-$ as an example \cite{LHCb:2019jta,LHCb:2019sus}. 
The experimental results of the LHCb collaboration indicate that there is a strong interfering effect between $\rho(770)^0$ and $f_0(500)$ in the phase space of the intermediate resonance $\rho(770)^0$ region \cite{Zhang:2013oqa}.
As is pointed out by the author of Ref. \cite{Cheng:2022ysn},the results of LHCb imply strongly that the imaginary part of the interfering terms $\textrm{Im}(c_-^{*\rho}c_-^{\sigma}+c_-^{\rho}c_-^{*\sigma})$  
gives large contribution to the regional CPAs \cite{Cheng:2022ysn}. 
Based on the data extracted from Fig. 12 of Ref. \cite{LHCb:2019sus}, we obtain the FB-CPA of each bin distributed in the invariant mass of the $\pi^+\pi^-$ pair with lower invariant mass $m_{\text{low}}$, as shown in Fig. \ref{fig:FBCPA3pi}. 
As can be inferred from Eq. (\ref{eq:FBCPA}), FB-CPA of each bin is defined as 
\begin{equation}
  A_{CP,k}^{FB}= \frac{(N_{B^-}-N_{B^+})_{{\cos\theta_{\text{hel}}>0},k}-(N_{B^-}-N_{B^+})_{{\cos\theta_{\text{hel}}<0},k}}{ (N_{B^-}+N_{B^+})_{{\cos\theta_{\text{hel}}>0},k}+(N_{B^-}+N_{B^+})_{{\cos\theta_{\text{hel}}<0},k}},
\end{equation}
where $k$ is the bin index.
From Fig. \ref{fig:FBCPA3pi} one can see that $A_{CP}^{FB}$ changes rapidly around the $\rho(770)^0$ resonance.
For the eight bins around the $\rho(770)^0$ resonance which are highlighted with grey shaded areas in Fig. \ref{fig:FBCPA3pi}, the averaged FB-CPA is obtained to be 
\begin{equation}\label{eq:AFBCPavebin}
A_{CP}^{FB, \text{ave}}=\frac{\sum_{k=8}^{15}[(N_{B^-}-N_{B^+})_{{\cos\theta_{\text{hel}}>0},k}-(N_{B^-}-N_{B^+})_{{\cos\theta_{\text{hel}}<0},k}]}{ \sum_{k=8}^{15} [(N_{B^-}+N_{B^+})_{{\cos\theta_{\text{hel}}>0},k}+(N_{B^-}+N_{B^+})_{{\cos\theta_{\text{hel}}<0},k}]}=(0.8\pm1.0)\%.
\end{equation}
This result shows clearly that while the FB-CPA of each bins around the $\rho(770)^0$ resonance are relatively large, there is a strong cancellation when combining to obtain the averaged one. 
On the other hand, by introducing a sign factor $\text{sgn}(s-{m_2^\prime}^2)$ for $m_2^\prime$ close to the mass of the $\rho(770)^0$ resonance, one can avoid the cancellation \footnote{It will not make much difference if we include more bins. For example, if bin 7 and bin 16 are included, Eqs. (\ref{eq:AFBCPavebin}) and (\ref{eq:AFBCPImbin}) become $A_{CP}^{FB, \text{ave}}=(-1.1\pm0.9)\%$ and $A_{CP}^{FB, \Im}=(12.8\pm0.9)\%$, respectively.}:   
 \begin{equation}\label{eq:AFBCPImbin}
 A_{CP}^{FB, \Im}=\frac{\left(\sum_{k=12}^{15}-\sum_{k=8}^{11}\right)\left[(N_{B^-}-N_{B^+})_{ {\cos\theta_{\text{hel}}>0},k}-(N_{B^-}-N_{B^+})_{{\cos\theta_{\text{hel}}<0},k}\right]}{ \sum_{k=8}^{15} \left[(N_{B^-}+N_{B^+})_{{\cos\theta_{\text{hel}}>0},k}+(N_{B^-}+N_{B^+})_{{\cos\theta_{\text{hel}}<0},k}\right]}=(13.2\pm1.0)\%.
 \end{equation}

The analysis of this paper is especially helpful for CPV searching in baryon decay processes.
Unlike the bottom and charm meson decays, the bottom and charmed baryon decay processes are quite limited by the substantial lower statistics.
It is impossible to see the interfering effect as clear as that in the $B$ meson decay processes such as $B^\pm\to \pi^\pm\pi^+\pi^-$.
As a consequence, a measurement of decay-angular distribution CPV observables such as the aforementioned $A_{CP}^{FB}$ and $A_{CP}^{(l)}$ in the interfering region only reveal the contribution of the term $\Re \left({\mathcal{A}_1\mathcal{A}_{2}^\ast}\right)$.
The contribution of the term $\Im \left({\mathcal{A}_1\mathcal{A}_{2}^\ast}\right)$ to CPV will simply be missed.
In order to search for CPV effect caused by the term $\Im \left({\mathcal{A}_1\mathcal{A}_{2}^\ast}\right)$ in bottom and charmed baryon decay processes, which has never been raised up explicitly in the literature, we strongly suggest to measure the observables such as $A_{CP}^{FB,\Im}$ and $A_{CP}^{(l),\Im}$.

\begin{acknowledgments}
This work was supported by National Natural Science Foundation of China under Grants Nos. 12475096, 12405115, 12192261, Natural Science Foundation of
Hunan Province under Grants No. 2022JJ30483, and Scientific Research Fund of Hunan Provincial Education Department under Grants No. 22A0319.
\end{acknowledgments}

\bibliography{zzhbib}

\begin{thebibliography}{38}%
\makeatletter
\providecommand \@ifxundefined [1]{%
 \@ifx{#1\undefined}
}%
\providecommand \@ifnum [1]{%
 \ifnum #1\expandafter \@firstoftwo
 \else \expandafter \@secondoftwo
 \fi
}%
\providecommand \@ifx [1]{%
 \ifx #1\expandafter \@firstoftwo
 \else \expandafter \@secondoftwo
 \fi
}%
\providecommand \natexlab [1]{#1}%
\providecommand \enquote  [1]{``#1''}%
\providecommand \bibnamefont  [1]{#1}%
\providecommand \bibfnamefont [1]{#1}%
\providecommand \citenamefont [1]{#1}%
\providecommand \href@noop [0]{\@secondoftwo}%
\providecommand \href [0]{\begingroup \@sanitize@url \@href}%
\providecommand \@href[1]{\@@startlink{#1}\@@href}%
\providecommand \@@href[1]{\endgroup#1\@@endlink}%
\providecommand \@sanitize@url [0]{\catcode `\\12\catcode `\$12\catcode
  `\&12\catcode `\#12\catcode `\^12\catcode `\_12\catcode `\%12\relax}%
\providecommand \@@startlink[1]{}%
\providecommand \@@endlink[0]{}%
\providecommand \url  [0]{\begingroup\@sanitize@url \@url }%
\providecommand \@url [1]{\endgroup\@href {#1}{\urlprefix }}%
\providecommand \urlprefix  [0]{URL }%
\providecommand \Eprint [0]{\href }%
\providecommand \doibase [0]{https://doi.org/}%
\providecommand \selectlanguage [0]{\@gobble}%
\providecommand \bibinfo  [0]{\@secondoftwo}%
\providecommand \bibfield  [0]{\@secondoftwo}%
\providecommand \translation [1]{[#1]}%
\providecommand \BibitemOpen [0]{}%
\providecommand \bibitemStop [0]{}%
\providecommand \bibitemNoStop [0]{.\EOS\space}%
\providecommand \EOS [0]{\spacefactor3000\relax}%
\providecommand \BibitemShut  [1]{\csname bibitem#1\endcsname}%
\let\auto@bib@innerbib\@empty
\bibitem [{\citenamefont {Christenson}\ \emph {et~al.}(1964)\citenamefont
  {Christenson}, \citenamefont {Cronin}, \citenamefont {Fitch},\ and\
  \citenamefont {Turlay}}]{Christenson:1964fg}%
  \BibitemOpen
  \bibfield  {author} {\bibinfo {author} {\bibfnamefont {J.~H.}\ \bibnamefont
  {Christenson}}, \bibinfo {author} {\bibfnamefont {J.~W.}\ \bibnamefont
  {Cronin}}, \bibinfo {author} {\bibfnamefont {V.~L.}\ \bibnamefont {Fitch}},\
  and\ \bibinfo {author} {\bibfnamefont {R.}~\bibnamefont {Turlay}},\
  }\bibfield  {title} {\bibinfo {title} {{Evidence for the $2\pi$ Decay of the
  $K_2^0$ Meson}},\ }\href {https://doi.org/10.1103/PhysRevLett.13.138}
  {\bibfield  {journal} {\bibinfo  {journal} {Phys. Rev. Lett.}\ }\textbf
  {\bibinfo {volume} {13}},\ \bibinfo {pages} {138} (\bibinfo {year}
  {1964})}\BibitemShut {NoStop}%
\bibitem [{\citenamefont {Aaij}\ \emph
  {et~al.}(2019{\natexlab{a}})\citenamefont {Aaij} \emph
  {et~al.}}]{LHCb:2019hro}%
  \BibitemOpen
  \bibfield  {author} {\bibinfo {author} {\bibfnamefont {R.}~\bibnamefont
  {Aaij}} \emph {et~al.} (\bibinfo {collaboration} {LHCb}),\ }\bibfield
  {title} {\bibinfo {title} {{Observation of CP Violation in Charm Decays}},\
  }\href {https://doi.org/10.1103/PhysRevLett.122.211803} {\bibfield  {journal}
  {\bibinfo  {journal} {Phys. Rev. Lett.}\ }\textbf {\bibinfo {volume} {122}},\
  \bibinfo {pages} {211803} (\bibinfo {year} {2019}{\natexlab{a}})},\ \Eprint
  {https://arxiv.org/abs/1903.08726} {arXiv:1903.08726 [hep-ex]} \BibitemShut
  {NoStop}%
\bibitem [{\citenamefont {Aubert}\ \emph {et~al.}(2001)\citenamefont {Aubert}
  \emph {et~al.}}]{BaBar:2001ags}%
  \BibitemOpen
  \bibfield  {author} {\bibinfo {author} {\bibfnamefont {B.}~\bibnamefont
  {Aubert}} \emph {et~al.} (\bibinfo {collaboration} {BaBar}),\ }\bibfield
  {title} {\bibinfo {title} {{Measurement of CP violating asymmetries in $B^0$
  decays to CP eigenstates}},\ }\href
  {https://doi.org/10.1103/PhysRevLett.86.2515} {\bibfield  {journal} {\bibinfo
   {journal} {Phys. Rev. Lett.}\ }\textbf {\bibinfo {volume} {86}},\ \bibinfo
  {pages} {2515} (\bibinfo {year} {2001})},\ \Eprint
  {https://arxiv.org/abs/hep-ex/0102030} {arXiv:hep-ex/0102030} \BibitemShut
  {NoStop}%
\bibitem [{\citenamefont {Abe}\ \emph {et~al.}(2001)\citenamefont {Abe} \emph
  {et~al.}}]{Belle:2001zzw}%
  \BibitemOpen
  \bibfield  {author} {\bibinfo {author} {\bibfnamefont {K.}~\bibnamefont
  {Abe}} \emph {et~al.} (\bibinfo {collaboration} {Belle}),\ }\bibfield
  {title} {\bibinfo {title} {{Observation of large CP violation in the neutral
  $B$ meson system}},\ }\href {https://doi.org/10.1103/PhysRevLett.87.091802}
  {\bibfield  {journal} {\bibinfo  {journal} {Phys. Rev. Lett.}\ }\textbf
  {\bibinfo {volume} {87}},\ \bibinfo {pages} {091802} (\bibinfo {year}
  {2001})},\ \Eprint {https://arxiv.org/abs/hep-ex/0107061}
  {arXiv:hep-ex/0107061} \BibitemShut {NoStop}%
\bibitem [{\citenamefont {Aaij}\ \emph {et~al.}(2013)\citenamefont {Aaij} \emph
  {et~al.}}]{LHCb:2013syl}%
  \BibitemOpen
  \bibfield  {author} {\bibinfo {author} {\bibfnamefont {R.}~\bibnamefont
  {Aaij}} \emph {et~al.} (\bibinfo {collaboration} {LHCb}),\ }\bibfield
  {title} {\bibinfo {title} {{First observation of $CP$ violation in the decays
  of $B^0_s$ mesons}},\ }\href {https://doi.org/10.1103/PhysRevLett.110.221601}
  {\bibfield  {journal} {\bibinfo  {journal} {Phys. Rev. Lett.}\ }\textbf
  {\bibinfo {volume} {110}},\ \bibinfo {pages} {221601} (\bibinfo {year}
  {2013})},\ \Eprint {https://arxiv.org/abs/1304.6173} {arXiv:1304.6173
  [hep-ex]} \BibitemShut {NoStop}%
\bibitem [{\citenamefont {Workman}(2022)}]{Workman:2022ynf}%
  \BibitemOpen
  \bibfield  {author} {\bibinfo {author} {\bibfnamefont {R.~L.}\ \bibnamefont
  {Workman}} (\bibinfo {collaboration} {Particle Data Group}),\ }\bibfield
  {title} {\bibinfo {title} {{Review of Particle Physics}},\ }\href
  {https://doi.org/10.1093/ptep/ptac097} {\bibfield  {journal} {\bibinfo
  {journal} {PTEP}\ }\textbf {\bibinfo {volume} {2022}},\ \bibinfo {pages}
  {083C01} (\bibinfo {year} {2022})}\BibitemShut {NoStop}%
\bibitem [{\citenamefont {Donoghue}\ \emph {et~al.}(1987)\citenamefont
  {Donoghue}, \citenamefont {Holstein},\ and\ \citenamefont
  {Valencia}}]{Donoghue:1987wu}%
  \BibitemOpen
  \bibfield  {author} {\bibinfo {author} {\bibfnamefont {J.~F.}\ \bibnamefont
  {Donoghue}}, \bibinfo {author} {\bibfnamefont {B.~R.}\ \bibnamefont
  {Holstein}},\ and\ \bibinfo {author} {\bibfnamefont {G.}~\bibnamefont
  {Valencia}},\ }\bibfield  {title} {\bibinfo {title} {{Survey of Present and
  Future Tests of {CP} Violation}},\ }\href
  {https://doi.org/10.1142/S0217751X87000144} {\bibfield  {journal} {\bibinfo
  {journal} {Int. J. Mod. Phys. A}\ }\textbf {\bibinfo {volume} {2}},\ \bibinfo
  {pages} {319} (\bibinfo {year} {1987})}\BibitemShut {NoStop}%
\bibitem [{\citenamefont {Valencia}(1989)}]{Valencia:1988it}%
  \BibitemOpen
  \bibfield  {author} {\bibinfo {author} {\bibfnamefont {G.}~\bibnamefont
  {Valencia}},\ }\bibfield  {title} {\bibinfo {title} {{Angular Correlations in
  the Decay $B \to V V$ and {CP} Violation}},\ }\href
  {https://doi.org/10.1103/PhysRevD.39.3339} {\bibfield  {journal} {\bibinfo
  {journal} {Phys. Rev. D}\ }\textbf {\bibinfo {volume} {39}},\ \bibinfo
  {pages} {3339} (\bibinfo {year} {1989})}\BibitemShut {NoStop}%
\bibitem [{\citenamefont {Dunietz}\ \emph {et~al.}(1991)\citenamefont
  {Dunietz}, \citenamefont {Quinn}, \citenamefont {Snyder}, \citenamefont
  {Toki},\ and\ \citenamefont {Lipkin}}]{Dunietz:1990cj}%
  \BibitemOpen
  \bibfield  {author} {\bibinfo {author} {\bibfnamefont {I.}~\bibnamefont
  {Dunietz}}, \bibinfo {author} {\bibfnamefont {H.~R.}\ \bibnamefont {Quinn}},
  \bibinfo {author} {\bibfnamefont {A.}~\bibnamefont {Snyder}}, \bibinfo
  {author} {\bibfnamefont {W.}~\bibnamefont {Toki}},\ and\ \bibinfo {author}
  {\bibfnamefont {H.~J.}\ \bibnamefont {Lipkin}},\ }\bibfield  {title}
  {\bibinfo {title} {{How to extract CP violating asymmetries from angular
  correlations}},\ }\href {https://doi.org/10.1103/PhysRevD.43.2193} {\bibfield
   {journal} {\bibinfo  {journal} {Phys. Rev. D}\ }\textbf {\bibinfo {volume}
  {43}},\ \bibinfo {pages} {2193} (\bibinfo {year} {1991})}\BibitemShut
  {NoStop}%
\bibitem [{\citenamefont {Golowich}\ and\ \citenamefont
  {Valencia}(1989)}]{Golowich:1988ig}%
  \BibitemOpen
  \bibfield  {author} {\bibinfo {author} {\bibfnamefont {E.}~\bibnamefont
  {Golowich}}\ and\ \bibinfo {author} {\bibfnamefont {G.}~\bibnamefont
  {Valencia}},\ }\bibfield  {title} {\bibinfo {title} {{Triple Product
  Correlations in Semileptonic $B^\pm$ Decays}},\ }\href
  {https://doi.org/10.1103/PhysRevD.40.112} {\bibfield  {journal} {\bibinfo
  {journal} {Phys. Rev. D}\ }\textbf {\bibinfo {volume} {40}},\ \bibinfo
  {pages} {112} (\bibinfo {year} {1989})}\BibitemShut {NoStop}%
\bibitem [{\citenamefont {Kayser}(1990)}]{Kayser:1989vw}%
  \BibitemOpen
  \bibfield  {author} {\bibinfo {author} {\bibfnamefont {B.}~\bibnamefont
  {Kayser}},\ }\bibfield  {title} {\bibinfo {title} {{Kinematically Nontrivial
  {CP} Violation in Beauty Decay}},\ }\href
  {https://doi.org/10.1016/0920-5632(90)90113-9} {\bibfield  {journal}
  {\bibinfo  {journal} {Nucl. Phys. B Proc. Suppl.}\ }\textbf {\bibinfo
  {volume} {13}},\ \bibinfo {pages} {487} (\bibinfo {year} {1990})}\BibitemShut
  {NoStop}%
\bibitem [{\citenamefont {Bensalem}\ and\ \citenamefont
  {London}(2001)}]{Bensalem:2000hq}%
  \BibitemOpen
  \bibfield  {author} {\bibinfo {author} {\bibfnamefont {W.}~\bibnamefont
  {Bensalem}}\ and\ \bibinfo {author} {\bibfnamefont {D.}~\bibnamefont
  {London}},\ }\bibfield  {title} {\bibinfo {title} {{$T$ odd triple product
  correlations in hadronic $b$ decays}},\ }\href
  {https://doi.org/10.1103/PhysRevD.64.116003} {\bibfield  {journal} {\bibinfo
  {journal} {Phys. Rev. D}\ }\textbf {\bibinfo {volume} {64}},\ \bibinfo
  {pages} {116003} (\bibinfo {year} {2001})},\ \Eprint
  {https://arxiv.org/abs/hep-ph/0005018} {arXiv:hep-ph/0005018} \BibitemShut
  {NoStop}%
\bibitem [{\citenamefont {Bensalem}\ \emph
  {et~al.}(2002{\natexlab{a}})\citenamefont {Bensalem}, \citenamefont {Datta},\
  and\ \citenamefont {London}}]{Bensalem:2002pz}%
  \BibitemOpen
  \bibfield  {author} {\bibinfo {author} {\bibfnamefont {W.}~\bibnamefont
  {Bensalem}}, \bibinfo {author} {\bibfnamefont {A.}~\bibnamefont {Datta}},\
  and\ \bibinfo {author} {\bibfnamefont {D.}~\bibnamefont {London}},\
  }\bibfield  {title} {\bibinfo {title} {{T violating triple product
  correlations in charmless Lambda(b) decays}},\ }\href
  {https://doi.org/10.1016/S0370-2693(02)02028-2} {\bibfield  {journal}
  {\bibinfo  {journal} {Phys. Lett. B}\ }\textbf {\bibinfo {volume} {538}},\
  \bibinfo {pages} {309} (\bibinfo {year} {2002}{\natexlab{a}})},\ \Eprint
  {https://arxiv.org/abs/hep-ph/0205009} {arXiv:hep-ph/0205009} \BibitemShut
  {NoStop}%
\bibitem [{\citenamefont {Bensalem}\ \emph
  {et~al.}(2002{\natexlab{b}})\citenamefont {Bensalem}, \citenamefont {Datta},\
  and\ \citenamefont {London}}]{Bensalem:2002ys}%
  \BibitemOpen
  \bibfield  {author} {\bibinfo {author} {\bibfnamefont {W.}~\bibnamefont
  {Bensalem}}, \bibinfo {author} {\bibfnamefont {A.}~\bibnamefont {Datta}},\
  and\ \bibinfo {author} {\bibfnamefont {D.}~\bibnamefont {London}},\
  }\bibfield  {title} {\bibinfo {title} {{New physics effects on triple product
  correlations in Lambda(b) decays}},\ }\href
  {https://doi.org/10.1103/PhysRevD.66.094004} {\bibfield  {journal} {\bibinfo
  {journal} {Phys. Rev. D}\ }\textbf {\bibinfo {volume} {66}},\ \bibinfo
  {pages} {094004} (\bibinfo {year} {2002}{\natexlab{b}})},\ \Eprint
  {https://arxiv.org/abs/hep-ph/0208054} {arXiv:hep-ph/0208054} \BibitemShut
  {NoStop}%
\bibitem [{\citenamefont {Durieux}\ and\ \citenamefont
  {Grossman}(2015)}]{Durieux:2015zwa}%
  \BibitemOpen
  \bibfield  {author} {\bibinfo {author} {\bibfnamefont {G.}~\bibnamefont
  {Durieux}}\ and\ \bibinfo {author} {\bibfnamefont {Y.}~\bibnamefont
  {Grossman}},\ }\bibfield  {title} {\bibinfo {title} {{Probing CP violation
  systematically in differential distributions}},\ }\href
  {https://doi.org/10.1103/PhysRevD.92.076013} {\bibfield  {journal} {\bibinfo
  {journal} {Phys. Rev. D}\ }\textbf {\bibinfo {volume} {92}},\ \bibinfo
  {pages} {076013} (\bibinfo {year} {2015})},\ \Eprint
  {https://arxiv.org/abs/1508.03054} {arXiv:1508.03054 [hep-ph]} \BibitemShut
  {NoStop}%
\bibitem [{\citenamefont {Gronau}\ and\ \citenamefont
  {Rosner}(2015)}]{Gronau:2015gha}%
  \BibitemOpen
  \bibfield  {author} {\bibinfo {author} {\bibfnamefont {M.}~\bibnamefont
  {Gronau}}\ and\ \bibinfo {author} {\bibfnamefont {J.~L.}\ \bibnamefont
  {Rosner}},\ }\bibfield  {title} {\bibinfo {title} {{Triple product
  asymmmetries in $\Lambda_b$ and $\Xi_b$ decays}},\ }\href
  {https://doi.org/10.1016/j.physletb.2015.07.060} {\bibfield  {journal}
  {\bibinfo  {journal} {Phys. Lett. B}\ }\textbf {\bibinfo {volume} {749}},\
  \bibinfo {pages} {104} (\bibinfo {year} {2015})},\ \Eprint
  {https://arxiv.org/abs/1506.01346} {arXiv:1506.01346 [hep-ph]} \BibitemShut
  {NoStop}%
\bibitem [{\citenamefont {Shi}\ \emph {et~al.}(2019)\citenamefont {Shi},
  \citenamefont {Kang}, \citenamefont {Bigi}, \citenamefont {Wang},\ and\
  \citenamefont {Peng}}]{Shi:2019vus}%
  \BibitemOpen
  \bibfield  {author} {\bibinfo {author} {\bibfnamefont {X.-D.}\ \bibnamefont
  {Shi}}, \bibinfo {author} {\bibfnamefont {X.-W.}\ \bibnamefont {Kang}},
  \bibinfo {author} {\bibfnamefont {I.}~\bibnamefont {Bigi}}, \bibinfo {author}
  {\bibfnamefont {W.-P.}\ \bibnamefont {Wang}},\ and\ \bibinfo {author}
  {\bibfnamefont {H.-P.}\ \bibnamefont {Peng}},\ }\bibfield  {title} {\bibinfo
  {title} {{Prospects for CP and P violation in $\Lambda_{c}^+$ decays at Super
  Tau Charm Facility}},\ }\href {https://doi.org/10.1103/PhysRevD.100.113002}
  {\bibfield  {journal} {\bibinfo  {journal} {Phys. Rev. D}\ }\textbf {\bibinfo
  {volume} {100}},\ \bibinfo {pages} {113002} (\bibinfo {year} {2019})},\
  \Eprint {https://arxiv.org/abs/1904.12415} {arXiv:1904.12415 [hep-ph]}
  \BibitemShut {NoStop}%
\bibitem [{\citenamefont {Wang}\ \emph {et~al.}(2022)\citenamefont {Wang},
  \citenamefont {Qin},\ and\ \citenamefont {Yu}}]{Wang:2022fih}%
  \BibitemOpen
  \bibfield  {author} {\bibinfo {author} {\bibfnamefont {J.-P.}\ \bibnamefont
  {Wang}}, \bibinfo {author} {\bibfnamefont {Q.}~\bibnamefont {Qin}},\ and\
  \bibinfo {author} {\bibfnamefont {F.-S.}\ \bibnamefont {Yu}},\ }\bibfield
  {title} {\bibinfo {title} {{CP violation induced by T-odd correlations and
  its baryonic application}},\ }\href@noop {} {\  (\bibinfo {year} {2022})},\
  \Eprint {https://arxiv.org/abs/2211.07332} {arXiv:2211.07332 [hep-ph]}
  \BibitemShut {NoStop}%
\bibitem [{\citenamefont {Itoh}\ \emph {et~al.}(2005)\citenamefont {Itoh} \emph
  {et~al.}}]{Belle:2005qtf}%
  \BibitemOpen
  \bibfield  {author} {\bibinfo {author} {\bibfnamefont {R.}~\bibnamefont
  {Itoh}} \emph {et~al.} (\bibinfo {collaboration} {Belle}),\ }\bibfield
  {title} {\bibinfo {title} {{Studies of CP violation in $B \to J/\psi K^\ast$
  decays}},\ }\href {https://doi.org/10.1103/PhysRevLett.95.091601} {\bibfield
  {journal} {\bibinfo  {journal} {Phys. Rev. Lett.}\ }\textbf {\bibinfo
  {volume} {95}},\ \bibinfo {pages} {091601} (\bibinfo {year} {2005})},\
  \Eprint {https://arxiv.org/abs/hep-ex/0504030} {arXiv:hep-ex/0504030}
  \BibitemShut {NoStop}%
\bibitem [{\citenamefont {del Amo~Sanchez}\ \emph {et~al.}(2010)\citenamefont
  {del Amo~Sanchez} \emph {et~al.}}]{BaBar:2010xrb}%
  \BibitemOpen
  \bibfield  {author} {\bibinfo {author} {\bibfnamefont {P.}~\bibnamefont {del
  Amo~Sanchez}} \emph {et~al.} (\bibinfo {collaboration} {BaBar}),\ }\bibfield
  {title} {\bibinfo {title} {{Search for CP violation using $T$-odd
  correlations in $D^0 \to K^+ K^- \pi^+ \pi^-$ decays}},\ }\href
  {https://doi.org/10.1103/PhysRevD.81.111103} {\bibfield  {journal} {\bibinfo
  {journal} {Phys. Rev. D}\ }\textbf {\bibinfo {volume} {81}},\ \bibinfo
  {pages} {111103} (\bibinfo {year} {2010})},\ \Eprint
  {https://arxiv.org/abs/1003.3397} {arXiv:1003.3397 [hep-ex]} \BibitemShut
  {NoStop}%
\bibitem [{\citenamefont {Aaij}\ \emph {et~al.}(2014)\citenamefont {Aaij} \emph
  {et~al.}}]{LHCb:2014djq}%
  \BibitemOpen
  \bibfield  {author} {\bibinfo {author} {\bibfnamefont {R.}~\bibnamefont
  {Aaij}} \emph {et~al.} (\bibinfo {collaboration} {LHCb}),\ }\bibfield
  {title} {\bibinfo {title} {{Search for $CP$ violation using $T$-odd
  correlations in $D^0 \to K^+K^-\pi^+\pi^-$ decays}},\ }\href
  {https://doi.org/10.1007/JHEP10(2014)005} {\bibfield  {journal} {\bibinfo
  {journal} {JHEP}\ }\textbf {\bibinfo {volume} {10}},\ \bibinfo {pages}
  {005}},\ \Eprint {https://arxiv.org/abs/1408.1299} {arXiv:1408.1299 [hep-ex]}
  \BibitemShut {NoStop}%
\bibitem [{\citenamefont {Aaij}\ \emph
  {et~al.}(2017{\natexlab{a}})\citenamefont {Aaij} \emph
  {et~al.}}]{LHCb:2016yco}%
  \BibitemOpen
  \bibfield  {author} {\bibinfo {author} {\bibfnamefont {R.}~\bibnamefont
  {Aaij}} \emph {et~al.} (\bibinfo {collaboration} {LHCb}),\ }\bibfield
  {title} {\bibinfo {title} {{Measurement of matter-antimatter differences in
  beauty baryon decays}},\ }\href {https://doi.org/10.1038/nphys4021}
  {\bibfield  {journal} {\bibinfo  {journal} {Nature Phys.}\ }\textbf {\bibinfo
  {volume} {13}},\ \bibinfo {pages} {391} (\bibinfo {year}
  {2017}{\natexlab{a}})},\ \Eprint {https://arxiv.org/abs/1609.05216}
  {arXiv:1609.05216 [hep-ex]} \BibitemShut {NoStop}%
\bibitem [{\citenamefont {Aaij}\ \emph
  {et~al.}(2017{\natexlab{b}})\citenamefont {Aaij} \emph
  {et~al.}}]{LHCb:2016qbq}%
  \BibitemOpen
  \bibfield  {author} {\bibinfo {author} {\bibfnamefont {R.}~\bibnamefont
  {Aaij}} \emph {et~al.} (\bibinfo {collaboration} {LHCb}),\ }\bibfield
  {title} {\bibinfo {title} {{Search for $CP$ violation in the phase space of
  $D^0\rightarrow\pi^+\pi^-\pi^+\pi^-$ decays}},\ }\href
  {https://doi.org/10.1016/j.physletb.2017.03.062} {\bibfield  {journal}
  {\bibinfo  {journal} {Phys. Lett. B}\ }\textbf {\bibinfo {volume} {769}},\
  \bibinfo {pages} {345} (\bibinfo {year} {2017}{\natexlab{b}})},\ \Eprint
  {https://arxiv.org/abs/1612.03207} {arXiv:1612.03207 [hep-ex]} \BibitemShut
  {NoStop}%
\bibitem [{\citenamefont {Prasanth}\ \emph {et~al.}(2017)\citenamefont
  {Prasanth} \emph {et~al.}}]{Belle:2017zvp}%
  \BibitemOpen
  \bibfield  {author} {\bibinfo {author} {\bibfnamefont {K.}~\bibnamefont
  {Prasanth}} \emph {et~al.} (\bibinfo {collaboration} {Belle}),\ }\bibfield
  {title} {\bibinfo {title} {{First measurement of ${T}$-odd moments in ${D^{0}
  \rightarrow K_{S}^{0} \pi^{+} \pi^{-} \pi^{0}}$ decays}},\ }\href
  {https://doi.org/10.1103/PhysRevD.95.091101} {\bibfield  {journal} {\bibinfo
  {journal} {Phys. Rev. D}\ }\textbf {\bibinfo {volume} {95}},\ \bibinfo
  {pages} {091101} (\bibinfo {year} {2017})},\ \Eprint
  {https://arxiv.org/abs/1703.05721} {arXiv:1703.05721 [hep-ex]} \BibitemShut
  {NoStop}%
\bibitem [{\citenamefont {Aaij}\ \emph {et~al.}(2018)\citenamefont {Aaij} \emph
  {et~al.}}]{LHCb:2018fpt}%
  \BibitemOpen
  \bibfield  {author} {\bibinfo {author} {\bibfnamefont {R.}~\bibnamefont
  {Aaij}} \emph {et~al.} (\bibinfo {collaboration} {LHCb}),\ }\bibfield
  {title} {\bibinfo {title} {{Search for CP violation using triple product
  asymmetries in $\Lambda^{0}_{b}\to pK^{-}\pi^{+}\pi^{-}$, $\Lambda^{0}_{b}\to
  pK^{-}K^{+}K^{-}$ and $\Xi^{0}_{b}\to pK^{-}K^{-}\pi^{+}$ decays}},\ }\href
  {https://doi.org/10.1007/JHEP08(2018)039} {\bibfield  {journal} {\bibinfo
  {journal} {JHEP}\ }\textbf {\bibinfo {volume} {08}},\ \bibinfo {pages}
  {039}},\ \Eprint {https://arxiv.org/abs/1805.03941} {arXiv:1805.03941
  [hep-ex]} \BibitemShut {NoStop}%
\bibitem [{\citenamefont {Kim}\ \emph {et~al.}(2019)\citenamefont {Kim} \emph
  {et~al.}}]{Belle:2018pcz}%
  \BibitemOpen
  \bibfield  {author} {\bibinfo {author} {\bibfnamefont {J.~B.}\ \bibnamefont
  {Kim}} \emph {et~al.} (\bibinfo {collaboration} {Belle}),\ }\bibfield
  {title} {\bibinfo {title} {{Search for $CP$ violation with kinematic
  asymmetries in the $D^0 \to K^+ K^- \pi^+ \pi^-$ decay}},\ }\href
  {https://doi.org/10.1103/PhysRevD.99.011104} {\bibfield  {journal} {\bibinfo
  {journal} {Phys. Rev. D}\ }\textbf {\bibinfo {volume} {99}},\ \bibinfo
  {pages} {011104} (\bibinfo {year} {2019})},\ \Eprint
  {https://arxiv.org/abs/1810.06457} {arXiv:1810.06457 [hep-ex]} \BibitemShut
  {NoStop}%
\bibitem [{\citenamefont {Aaij}\ \emph
  {et~al.}(2019{\natexlab{b}})\citenamefont {Aaij} \emph
  {et~al.}}]{LHCb:2018mzv}%
  \BibitemOpen
  \bibfield  {author} {\bibinfo {author} {\bibfnamefont {R.}~\bibnamefont
  {Aaij}} \emph {et~al.} (\bibinfo {collaboration} {LHCb}),\ }\bibfield
  {title} {\bibinfo {title} {{Search for $CP$ violation through an amplitude
  analysis of $D^0 \to K^+ K^- \pi^+ \pi^-$ decays}},\ }\href
  {https://doi.org/10.1007/JHEP02(2019)126} {\bibfield  {journal} {\bibinfo
  {journal} {JHEP}\ }\textbf {\bibinfo {volume} {02}},\ \bibinfo {pages}
  {126}},\ \Eprint {https://arxiv.org/abs/1811.08304} {arXiv:1811.08304
  [hep-ex]} \BibitemShut {NoStop}%
\bibitem [{\citenamefont {Aaij}\ \emph
  {et~al.}(2020{\natexlab{a}})\citenamefont {Aaij} \emph
  {et~al.}}]{LHCb:2019oke}%
  \BibitemOpen
  \bibfield  {author} {\bibinfo {author} {\bibfnamefont {R.}~\bibnamefont
  {Aaij}} \emph {et~al.} (\bibinfo {collaboration} {LHCb}),\ }\bibfield
  {title} {\bibinfo {title} {{Search for $CP$ violation and observation of $P$
  violation in $\Lambda_b^0 \to p \pi^- \pi^+ \pi^-$ decays}},\ }\href
  {https://doi.org/10.1103/PhysRevD.102.051101} {\bibfield  {journal} {\bibinfo
   {journal} {Phys. Rev. D}\ }\textbf {\bibinfo {volume} {102}},\ \bibinfo
  {pages} {051101} (\bibinfo {year} {2020}{\natexlab{a}})},\ \Eprint
  {https://arxiv.org/abs/1912.10741} {arXiv:1912.10741 [hep-ex]} \BibitemShut
  {NoStop}%
\bibitem [{\citenamefont {Moon}\ \emph {et~al.}(2023)\citenamefont {Moon} \emph
  {et~al.}}]{Belle:2023bzn}%
  \BibitemOpen
  \bibfield  {author} {\bibinfo {author} {\bibfnamefont {H.~K.}\ \bibnamefont
  {Moon}} \emph {et~al.} (\bibinfo {collaboration} {Belle}),\ }\bibfield
  {title} {\bibinfo {title} {{Search for CP violation in $D_{(s)}^+\to K^+
  K_{S}^{0}h^+h^-$ ($h=K,\pi$) decays and observation of the Cabibbo-suppressed
  decay $D_s^+\to K^+K^-K_S^0\pi^+$}},\ }\href
  {https://doi.org/10.1103/PhysRevD.108.L111102} {\bibfield  {journal}
  {\bibinfo  {journal} {Phys. Rev. D}\ }\textbf {\bibinfo {volume} {108}},\
  \bibinfo {pages} {L111102} (\bibinfo {year} {2023})},\ \Eprint
  {https://arxiv.org/abs/2305.11405} {arXiv:2305.11405 [hep-ex]} \BibitemShut
  {NoStop}%
\bibitem [{\citenamefont {Wei}\ and\ \citenamefont
  {Zhang}(2022)}]{Wei:2022zuf}%
  \BibitemOpen
  \bibfield  {author} {\bibinfo {author} {\bibfnamefont {Y.-R.}\ \bibnamefont
  {Wei}}\ and\ \bibinfo {author} {\bibfnamefont {Z.-H.}\ \bibnamefont
  {Zhang}},\ }\bibfield  {title} {\bibinfo {title} {{Forward-backward asymmetry
  induced CP asymmetry of
  B\ensuremath{\pm}\textrightarrow{}\ensuremath{\pi}\ensuremath{\pm}\ensuremath{\pi}+\ensuremath{\pi}-}},\
  }\href {https://doi.org/10.1103/PhysRevD.106.113002} {\bibfield  {journal}
  {\bibinfo  {journal} {Phys. Rev. D}\ }\textbf {\bibinfo {volume} {106}},\
  \bibinfo {pages} {113002} (\bibinfo {year} {2022})},\ \Eprint
  {https://arxiv.org/abs/2209.02348} {arXiv:2209.02348 [hep-ph]} \BibitemShut
  {NoStop}%
\bibitem [{\citenamefont {Zhang}(2021)}]{Zhang:2021zhr}%
  \BibitemOpen
  \bibfield  {author} {\bibinfo {author} {\bibfnamefont {Z.-H.}\ \bibnamefont
  {Zhang}},\ }\bibfield  {title} {\bibinfo {title} {{A novel observable for CP
  violation in multi-body decays and its application potential to charm and
  beauty meson decays}},\ }\href
  {https://doi.org/10.1016/j.physletb.2021.136537} {\bibfield  {journal}
  {\bibinfo  {journal} {Phys. Lett. B}\ }\textbf {\bibinfo {volume} {820}},\
  \bibinfo {pages} {136537} (\bibinfo {year} {2021})},\ \Eprint
  {https://arxiv.org/abs/2102.12263} {arXiv:2102.12263 [hep-ph]} \BibitemShut
  {NoStop}%
\bibitem [{\citenamefont {Zhang}\ and\ \citenamefont
  {Qi}(2023)}]{Zhang:2022emj}%
  \BibitemOpen
  \bibfield  {author} {\bibinfo {author} {\bibfnamefont {Z.-H.}\ \bibnamefont
  {Zhang}}\ and\ \bibinfo {author} {\bibfnamefont {J.-J.}\ \bibnamefont {Qi}},\
  }\bibfield  {title} {\bibinfo {title} {{Analysis of angular distribution
  asymmetries and the associated $C\!P$ asymmetries in three-body decays of
  bottom baryons}},\ }\href {https://doi.org/10.1140/epjc/s10052-023-11267-7}
  {\bibfield  {journal} {\bibinfo  {journal} {Eur. Phys. J. C}\ }\textbf
  {\bibinfo {volume} {83}},\ \bibinfo {pages} {133} (\bibinfo {year} {2023})},\
  \Eprint {https://arxiv.org/abs/2208.13411} {arXiv:2208.13411 [hep-ph]}
  \BibitemShut {NoStop}%
\bibitem [{\citenamefont {Zhao}\ \emph {et~al.}(2024)\citenamefont {Zhao},
  \citenamefont {Zhang},\ and\ \citenamefont {Guo}}]{Zhao:2024ren}%
  \BibitemOpen
  \bibfield  {author} {\bibinfo {author} {\bibfnamefont {Y.-J.}\ \bibnamefont
  {Zhao}}, \bibinfo {author} {\bibfnamefont {Z.-H.}\ \bibnamefont {Zhang}},\
  and\ \bibinfo {author} {\bibfnamefont {X.-H.}\ \bibnamefont {Guo}},\
  }\bibfield  {title} {\bibinfo {title} {{Decay-angular-distribution correlated
  CP violation in heavy hadron cascade decays}},\ }\href
  {https://doi.org/10.1103/PhysRevD.110.013007} {\bibfield  {journal} {\bibinfo
   {journal} {Phys. Rev. D}\ }\textbf {\bibinfo {volume} {110}},\ \bibinfo
  {pages} {013007} (\bibinfo {year} {2024})},\ \Eprint
  {https://arxiv.org/abs/2403.05011} {arXiv:2403.05011 [hep-ph]} \BibitemShut
  {NoStop}%
\bibitem [{\citenamefont {Yu}\ \emph {et~al.}(2024)\citenamefont {Yu},
  \citenamefont {Han}, \citenamefont {Li}, \citenamefont {Li}, \citenamefont
  {Xiao},\ and\ \citenamefont {Yu}}]{Yu:2024cjd}%
  \BibitemOpen
  \bibfield  {author} {\bibinfo {author} {\bibfnamefont {J.-X.}\ \bibnamefont
  {Yu}}, \bibinfo {author} {\bibfnamefont {J.-J.}\ \bibnamefont {Han}},
  \bibinfo {author} {\bibfnamefont {Y.}~\bibnamefont {Li}}, \bibinfo {author}
  {\bibfnamefont {H.-n.}\ \bibnamefont {Li}}, \bibinfo {author} {\bibfnamefont
  {Z.-J.}\ \bibnamefont {Xiao}},\ and\ \bibinfo {author} {\bibfnamefont
  {F.-S.}\ \bibnamefont {Yu}},\ }\bibfield  {title} {\bibinfo {title}
  {{Establishing CP violation in $b$-baryon decays}},\ }\href@noop {} {\
  (\bibinfo {year} {2024})},\ \Eprint {https://arxiv.org/abs/2409.02821}
  {arXiv:2409.02821 [hep-ph]} \BibitemShut {NoStop}%
\bibitem [{\citenamefont {Aaij}\ \emph
  {et~al.}(2020{\natexlab{b}})\citenamefont {Aaij} \emph
  {et~al.}}]{LHCb:2019jta}%
  \BibitemOpen
  \bibfield  {author} {\bibinfo {author} {\bibfnamefont {R.}~\bibnamefont
  {Aaij}} \emph {et~al.} (\bibinfo {collaboration} {LHCb}),\ }\bibfield
  {title} {\bibinfo {title} {{Observation of Several Sources of $CP$ Violation
  in $B^+ \to \pi^+ \pi^+ \pi^-$ Decays}},\ }\href
  {https://doi.org/10.1103/PhysRevLett.124.031801} {\bibfield  {journal}
  {\bibinfo  {journal} {Phys. Rev. Lett.}\ }\textbf {\bibinfo {volume} {124}},\
  \bibinfo {pages} {031801} (\bibinfo {year} {2020}{\natexlab{b}})},\ \Eprint
  {https://arxiv.org/abs/1909.05211} {arXiv:1909.05211 [hep-ex]} \BibitemShut
  {NoStop}%
\bibitem [{\citenamefont {Aaij}\ \emph
  {et~al.}(2020{\natexlab{c}})\citenamefont {Aaij} \emph
  {et~al.}}]{LHCb:2019sus}%
  \BibitemOpen
  \bibfield  {author} {\bibinfo {author} {\bibfnamefont {R.}~\bibnamefont
  {Aaij}} \emph {et~al.} (\bibinfo {collaboration} {LHCb}),\ }\bibfield
  {title} {\bibinfo {title} {{Amplitude analysis of the $B^+ \rightarrow
  \pi^+\pi^+\pi^-$ decay}},\ }\href
  {https://doi.org/10.1103/PhysRevD.101.012006} {\bibfield  {journal} {\bibinfo
   {journal} {Phys. Rev. D}\ }\textbf {\bibinfo {volume} {101}},\ \bibinfo
  {pages} {012006} (\bibinfo {year} {2020}{\natexlab{c}})},\ \Eprint
  {https://arxiv.org/abs/1909.05212} {arXiv:1909.05212 [hep-ex]} \BibitemShut
  {NoStop}%
\bibitem [{\citenamefont {Zhang}\ \emph {et~al.}(2013)\citenamefont {Zhang},
  \citenamefont {Guo},\ and\ \citenamefont {Yang}}]{Zhang:2013oqa}%
  \BibitemOpen
  \bibfield  {author} {\bibinfo {author} {\bibfnamefont {Z.-H.}\ \bibnamefont
  {Zhang}}, \bibinfo {author} {\bibfnamefont {X.-H.}\ \bibnamefont {Guo}},\
  and\ \bibinfo {author} {\bibfnamefont {Y.-D.}\ \bibnamefont {Yang}},\
  }\bibfield  {title} {\bibinfo {title} {{CP violation in $B^{\pm} \rightarrow
  \pi^{\pm}\pi^{+}\pi^{-}$ in the region with low invariant mass of one
  $\pi^{+}\pi^{-}$ pair}},\ }\href {https://doi.org/10.1103/PhysRevD.87.076007}
  {\bibfield  {journal} {\bibinfo  {journal} {Phys. Rev. D}\ }\textbf {\bibinfo
  {volume} {87}},\ \bibinfo {pages} {076007} (\bibinfo {year} {2013})},\
  \Eprint {https://arxiv.org/abs/1303.3676} {arXiv:1303.3676 [hep-ph]}
  \BibitemShut {NoStop}%
\bibitem [{\citenamefont {Cheng}(2022)}]{Cheng:2022ysn}%
  \BibitemOpen
  \bibfield  {author} {\bibinfo {author} {\bibfnamefont {H.-Y.}\ \bibnamefont
  {Cheng}},\ }\bibfield  {title} {\bibinfo {title} {{CP violation in the
  interference between $\rho(770)^0$ and S-wave in $B^\pm\to \pi^+\pi^-\pi^\pm$
  decays and its implication for CP asymmetry in $B^\pm\to\rho^0\pi^\pm$}},\
  }\href {https://doi.org/10.1103/PhysRevD.106.113004} {\bibfield  {journal}
  {\bibinfo  {journal} {Phys. Rev. D}\ }\textbf {\bibinfo {volume} {106}},\
  \bibinfo {pages} {113004} (\bibinfo {year} {2022})},\ \Eprint
  {https://arxiv.org/abs/2211.03965} {arXiv:2211.03965 [hep-ph]} \BibitemShut
  {NoStop}%
\end{thebibliography}%

\end{document}